   \documentclass[final,5p,times,twocolumn]{elsarticle}
    
     \usepackage{amsmath}
    \usepackage{amssymb}
    \usepackage{hyperref}
      \usepackage{cleveref}
    \hypersetup{colorlinks,linkcolor={black},citecolor={blue},urlcolor={blue}}
    \usepackage{lipsum}
    \usepackage{booktabs}

    \makeatletter
    \def\ps@pprintTitle{%
       \let\@oddhead\@empty
       \let\@evenhead\@empty
       \let\@oddfoot\@empty
       \let\@evenfoot\@empty
    }
    \makeatother
\begin{document}
    
    \begin{frontmatter}
    
    

    \title{Tailoring the Growth of $\beta$-Tungsten Using Substrate Bias and Its Effect on FMR-Driven Spin Pumping in $\beta$-W/Py Heterostructures}
    

  \author[one]{Abhay Singh Rajawat}
  \author[one]{Naim Ahmad}
  \author[one]{Risvana Nasril}
  \author[one]{Tasneem Sheikh}
  \affiliation[one]{organization={Department of Physics, Jamia Millia Islamia}, 
          addressline={New Delhi}, 
        postcode={110025}, 
          country={India}}

\author[two]{Mohammad Muhiuddin}
\affiliation[two]{organization={Department of Metallurgical and Materials Engineering, National Institute of Technology }, 
          addressline={Karnataka}, 
          postcode={575014}, 
          country={India}}

\author[third]{Savita Sahu}
\affiliation[third]{organization={CSIR-National Physical Laboratory}, 
          addressline={Dr K S Krishnan Marg}, 
          city={New Delhi},
          postcode={110012}, 
          country={India}}

\author[four]{Ashwani Gautam}
\affiliation[four]{organization={Department of Ceramic Engineering, Indian Institute of Technology (BHU)}, 
          addressline={Varanasi}, 
          postcode={221005}, 
          country={India}}

\author[five]{A Kumar}
\affiliation[five]{organization={School of Physical Sciences, Jawaharlal Nehru University}, 
          addressline={New Delhi}, 
          postcode={110067}, 
          country={India}}

\author[four]{Md. Imteyaz Ahmad}
\author[third]{G. A. Basheed}
\author[two]{Mohammad R. Rahman}

\author[one]{Waseem Akhtar\corref{cor1}}
\ead{makhtar5@jmi.ac.in}
\cortext[cor1]{Corresponding author}

\begin{abstract}
$\beta$-Tungsten ($\beta$-W), an A15 cubic phase of tungsten, exhibits a giant spin Hall angle compared to its bcc-phase $\alpha$-Tungsten ($\alpha$-W), making high-quality $\beta$-W films desirable for spintronic applications. We report the controlled growth of $\beta$-W films on SiO$_2$/Si substrates via DC sputtering, where substrate bias serves as a critical factor in stabilizing the $\beta$ phase by regulating the energy of deposited atoms. This approach enables the formation of $\beta$-W films over a wide thickness range. Additionally, we studied the spin pumping phenomena in different tungsten phases achieved through substrate bias. Ferromagnetic resonance measurements reveal an enhancement in the magnetic damping (\( \alpha_{\text{eff}} \)) for $\beta$-W/Py compared to $\alpha$-W/Py dominated film. The effective spin mixing conductance (\( g_{\text{eff}}^{\uparrow\downarrow} \)) is found to be higher in $\beta$-W/Py than in $\alpha$-W/Py, which is attributed to variations in the interface structures between these phases. Importantly, the use of substrate bias does not deteriorate the interface quality, underscoring its effectiveness.  These findings highlight the potential of substrate bias in thin-film engineering, paving the way for its advanced utilization in spintronic applications.
\end{abstract}




\end{frontmatter}




\section{Introduction}
\label{introduction}

Heavy metal (HM) thin films are critical building blocks in spintronics technology, providing the means for efficient spin manipulation and transport, which are essential for the development of next-generation electronic and computing devices\cite{1,2}. Owing to the large spin-orbit coupling, HMs are efficient in converting charge current into spin current\cite{3,4}. Moreover, in heterostructure with magnetic layers, HMs leads to 
generation of spin-orbit torques for magnetic switching
as well as induces interfacial Dzyaloshinskii-Moriya interaction (DMI) that stabilizes exotic spin textures such as chiral domain walls as well as Skyrmions in magnetic layers\cite{5,6,7,8,9}. 
Heavy metals like platinum (Pt), tungsten (W), palladium (Pd), and others have been extensively studied over the last few decades to understand their functionalities and potential applications in spintronics\cite{10,11,12,13,14,15}. 

Tungsten, as a heavy metal, exists in two major phases: $\alpha$-tungsten ($\alpha$-W), which has a body-centered cubic (BCC) crystal structure, and $\beta$-tungsten ($\beta$-W), characterized by an A15 crystal structure\cite{16,17,18}. The $\beta$-phase has attracted significant attention in spintronics due to its giant spin Hall effect\cite{11,12,19,20,21}. Despite its significant potential, achieving a pure $\beta$-phase tungsten thin film remains a considerable challenge. The precise mechanisms governing the formation of the $\beta$-phase are not yet fully understood, prompting extensive research into this area. Deposition parameters such as deposition rate, gas pressure, substrate temperature, and film thickness have been extensively studied for their influence on the formation of the $\beta$-W phase \cite{16,18,20,21,22,23,24,25,26,27,28,29,30,31}. One key observation is the incorporation of oxygen impurities during growth as a major factor in the formation of the $\beta$-phase. These impurities are believed to play a crucial role in stabilizing the $\beta$-phase crystal structure\cite{16,32,33}. Furthermore, it is noted that the phase transition from the $\beta$-phase to the $\alpha$-phase occurs around a critical thickness close to the spin diffusion length in tungsten, highlighting that in this way, the $\beta$-phase can only remain stable up to a few nanometers\cite{11,12,31,34}. Additionally, the use of oxygen is not ideal for the chamber, as it oxidizes other targets and creates water vapor.

\begin{figure*}[h]
  \centering  \includegraphics[width=0.9\textwidth]{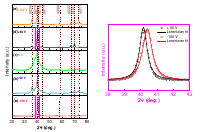}
  \caption{GIXRD pattern of sputter-deposited W films at substrate bias voltages of (a) -100 V, (b) -50 V, (c) 0 V, (d) +50 V and (e) +100 V. Dotted lines are guides to the eye, highlighting the crystallographic orientation of the $\alpha$ and $\beta$ phases of W. The zoomed-in image shows the 40° peak for +50 V and -100 V separately. }
  \label{fig:yourlabel}
\end{figure*}

In the current work, we are using substrate bias, which is less likely to introduce impurities and offers several other advantages, such as improved control over film properties, selective phase formation, and optimized growth kinetics\cite{18}. Here, we demonstrate a controlled phase transition (as seen in XRD) in films with thicknesses much larger than the spin diffusion length using substrate bias during DC sputtering. The surface morphology of these films has been analyzed using Field Emission Scanning Electron Microscopy (FESEM) and Atomic Force Microscopy (AFM). One-dimensional (1D) power spectral density (PSD) analysis has been performed to determine the RMS roughness (\( \sigma \)) and correlation length (\( \tau \)) of the samples deposited at different bias voltages. We further investigate the consistency of substrate bias effects at lower thicknesses. Our findings reveal that the results are consistent, with the film deposited with a +50 V bias being dominated by the $\beta$-W phase, whereas the film deposited with a -100 V bias is dominated by the $\alpha$-W phase. We perform magnetization measurements using the Magneto-Optic Kerr Effect (MOKE). The in-palne MOKE measurements suggest a complete in-plane easy axis of magnetization for the deposited bilayer samples. The Ferromagnetic Resonance (FMR) technique will be used to study the magnetization dynamics of the samples.

Here, in our study, we claim that substrate bias is helpful for stabilizing the $\beta$-W without the need of oxygen plasma. This phase, characterized by very high spin-orbit coupling (SOC), has wide applications in spintronics. The enhancement in  \(\alpha_{\text{eff}}\)  of $\beta$-W/Py thin films, measured using the FMR technique, indicates spin pumping. The large value of \( g_{\text{eff}}^{\uparrow\downarrow} \) for the $\beta$-W/Py heterostructure signifies the efficient transfer of spin current. These spin dynamic parameters differ in the different phases of W, which ultimately depend on the applied substrate bias.

\section{Experimental details}

We have grown a 2 series of tungsten films using DC magnetron sputtering on SiO$_2$ coated silicon (Si/SiO$_2$) substrates. The silicon substrates (SiO$_2$/Si) underwent a cleaning process involving ultrasonication for 10 minutes in acetone, isopropyl alcohol (IPA), and DI water, respectively followed by drying with nitrogen gas before being mounted on an electrically floated substrate holder in the vacuum chamber. A 2-inch cathode target of Tungsten (99.95\% purity),  procured from Testbourne, UK, was maintained at a distance of 10 cm from the substrate holder. The base pressure for the deposition was $6\times10^{-7}$ Torr. For sputtering, we obtain plasma with an Ar gas flow rate of 18 standard cubic centimetres per minute (sccm) controlled by a mass flow controller (MFC) maintaining a working pressure of 1 mTorr. Before starting the deposition, the W target was pre-sputtered for 5 mins to remove contamination from the target surface. 

In the first series, we deposited five tungsten films at substrate bias voltages of -100 V, -50 V, 0 V, +50 V, and +100 V, with the sputtering time fixed at 30 minutes. Thus, this series consists of thick tungsten films. In the second series, we grew two tungsten films at substrate bias voltages of +50 V and -100 V only, with a fixed thickness of 30 nm. All films were deposited under similar conditions except the applied substrate bias voltage. After optimizing the tungsten thin films, we fabricated bilayer structures by incorporating Py (80\% Ni and 20\% Fe) with tungsten (W) in two batches as listed in table 2. Batch 1 consisted of W(-100 V)(30 nm)/Py($x$)/SiO$_2$/Si, while Batch 2 consisted of W(+50 V)(30 nm)/Py($x$)/SiO$_2$/Si. These are referred to as $\alpha$-W(30)/Py($x$) and $\beta$-W(30)/Py($x$), respectively, where $x = 10$ nm and 20 nm. Therefore, we prepared a total of four bilayer samples: $\alpha$-W(30)/Py(10) and $\alpha$-W(30)/Py(20) in batch 1, as well as $\beta$-W(30)/Py(10) and $\beta$-W(30)/Py(20) in batch 2 .

We performed X-ray diffraction (XRD) for the structural analysis of the films. The surface morphology was analyzed using scanning electron microscopy (SEM) and atomic force microscopy (AFM). X-ray photoelectron spectroscopy (XPS) has been performed to analyze the chemical composition of W. The thickness and associated interface roughness of the bilayer structures were measured using X-ray reflectivity (XRR). The magnetic properties of the samples were studied using the magneto-optic Kerr effect (MOKE) and ferromagnetic resonance (FMR), which were integrated with a broadband coplanar waveguide (CPW) and a lock-in amplifier, covering a frequency range of 5–12 GHz.

\section{Results and discussion}
\subsection{\textbf{On-Demand Growth of $\beta$-W Film Using Substrate Bias}}

The crystal structure of the as-grown films of W under different substrate bias conditions was studied using X-ray diffraction. Figure 1 shows the Grazing Incidence X-ray Diffraction (GIXRD) pattern at grazing angle of incidence is 1$^\circ$ within the 2$\theta$ range of 20$^\circ$ to 80$^\circ$ acquired utilizing Cu K-$\alpha$ radiation ($\lambda$ = 1.54Å). 

At zero-bias (Figure 1(c)) a broad peak is observed around 2$\theta$ = 40$^\circ$ which indicates a lack of long-range crystalline order. As the negative bias of -50 V is applied during growth (Figure 1(b)), we observe a semi-crystalline film where the broad peak at 2$\theta$ = 40$^\circ$ starts to resolve into fine structures as well as some additional broad peaks appear around 2$\theta$ = 70$^\circ$. However, these peak positions confirm that the grown film at -50 V bias has a mixed phase of $\alpha$ and $\beta$-W.  Further increasing the negative bias voltage to -100 V (Figure 1(a)), we observe sharp GIXRD peaks highlighting the crystalline nature of the film with peak positions corresponding to  $\alpha$-W and indicating a planer orientation along $\alpha$-(110) \cite{18,35}. Based on the peak positions,  a lattice constant of 3.155 Å is obtained \cite{16}. Now, with a positive bias of +50 V (Figure 1(d)), a crystalline film is formed, with the peak positions matching the $\beta$-W phase and a lattice constant of 5.084 Å\cite{16,18,35}. This demonstrates a controlled growth of $\beta$-W film without using the oxygen plasma. 
As the bias voltage is increased to +100 V (Figure 1(e)), the film is crystalline but shows a mixed phase of $\alpha$ and $\beta$-W.The zoomed-in image, showing the 40° peak for +50 V and -100 V, clearly illustrates the shift between the $\beta$- and  $\alpha$-W phases. It can be concluded that at +50 V, the tungsten film is predominantly in the $\beta$ phase, while at -100 V, it is dominated by the $\alpha$ phase. Thus, the GIXRD data demonstrate that the resulting phase of the tungsten film strongly depends on the substrate bias voltage. The substrate bias besides controlling the crystallinity, can also be used for on-demand growth of  $\beta$ phase dominated W film.

\begin{figure}[h]
    \centering    \includegraphics[width=0.5\textwidth]{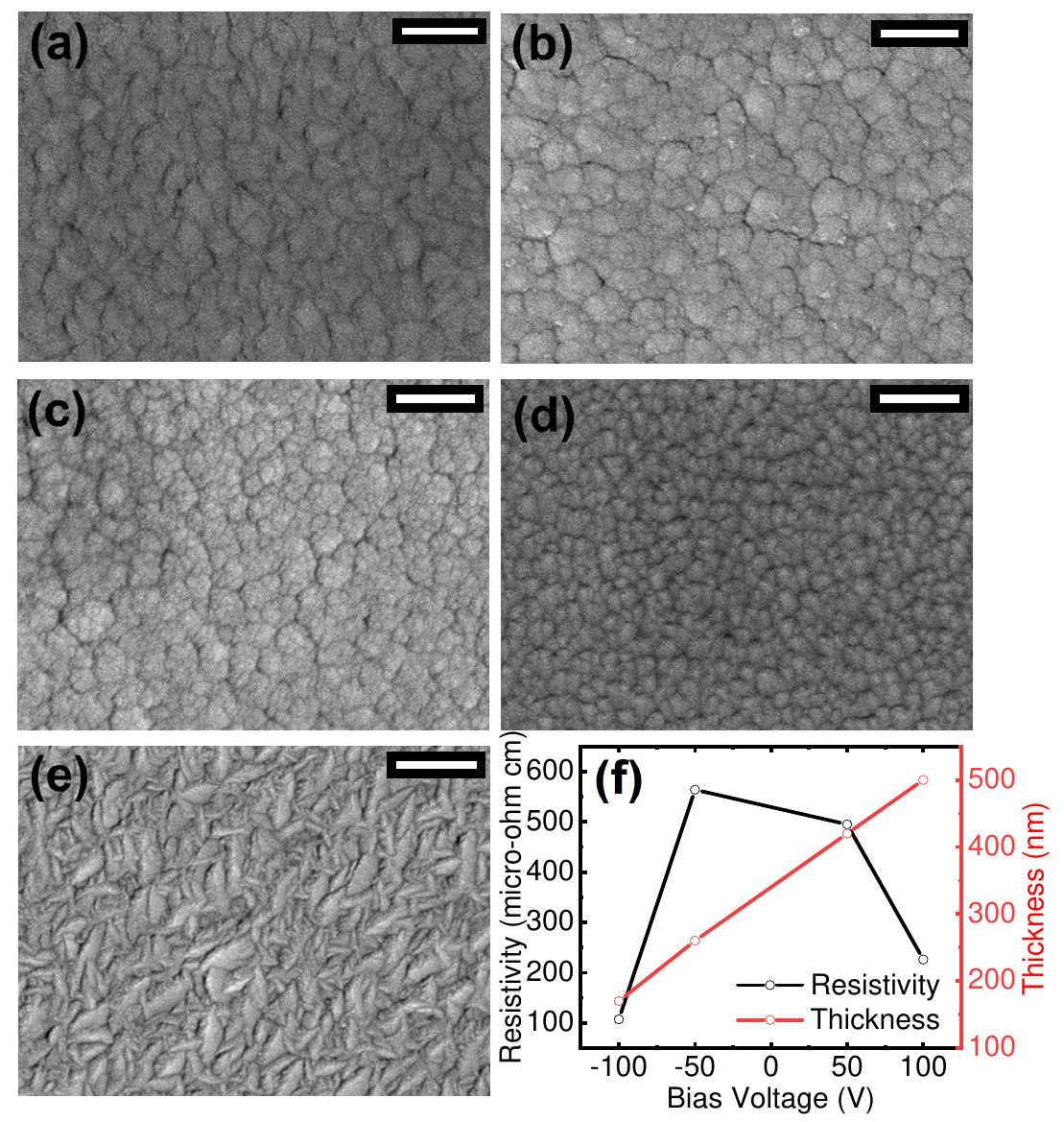}
    \caption{FESEM images of sputtered W films at substrate bias voltages of  (scale bar, 300 nm) (a) -100 V, (b) -50 V, (c) 0 V, (d) +50 V and (e) +100 V. (f) Resistivity and thickness of the deposited films as a function of bias voltage.}
    \label{fig:my_figure}
\end{figure}

We further used Field Emmission Scanning Electron Microscope (FESEM) to study the surface morphology of the deposited W film on SiO$_2$/Si substrate under different substrate bias conditions (Figure 2). At zero (Figure 2(c)) as well as at -50 V (Figure 2(b)) bias voltage, cauliflower-like structure is observed. As the negative bias increases to -100 V (Figure 2(a)), the morphology evolves into polyhedron grains of dimensions $\approx$100 nm. For positive bias of +50 V (Figure 2(d)), the uniform polyhedron grains of dimensions $\approx$ 40 nm are observed. However, the films deposited at +100 V (Figure 2(e)) exhibited needle-like structure characterised by grain sizes of nearly 100 nm. SEM images shows that $\beta$ dominated W film grown at +50 V substrate bias has the smallest grain size \cite{17}. 

\begin{figure*}[h]
  \centering  \includegraphics[width=01.0\textwidth]{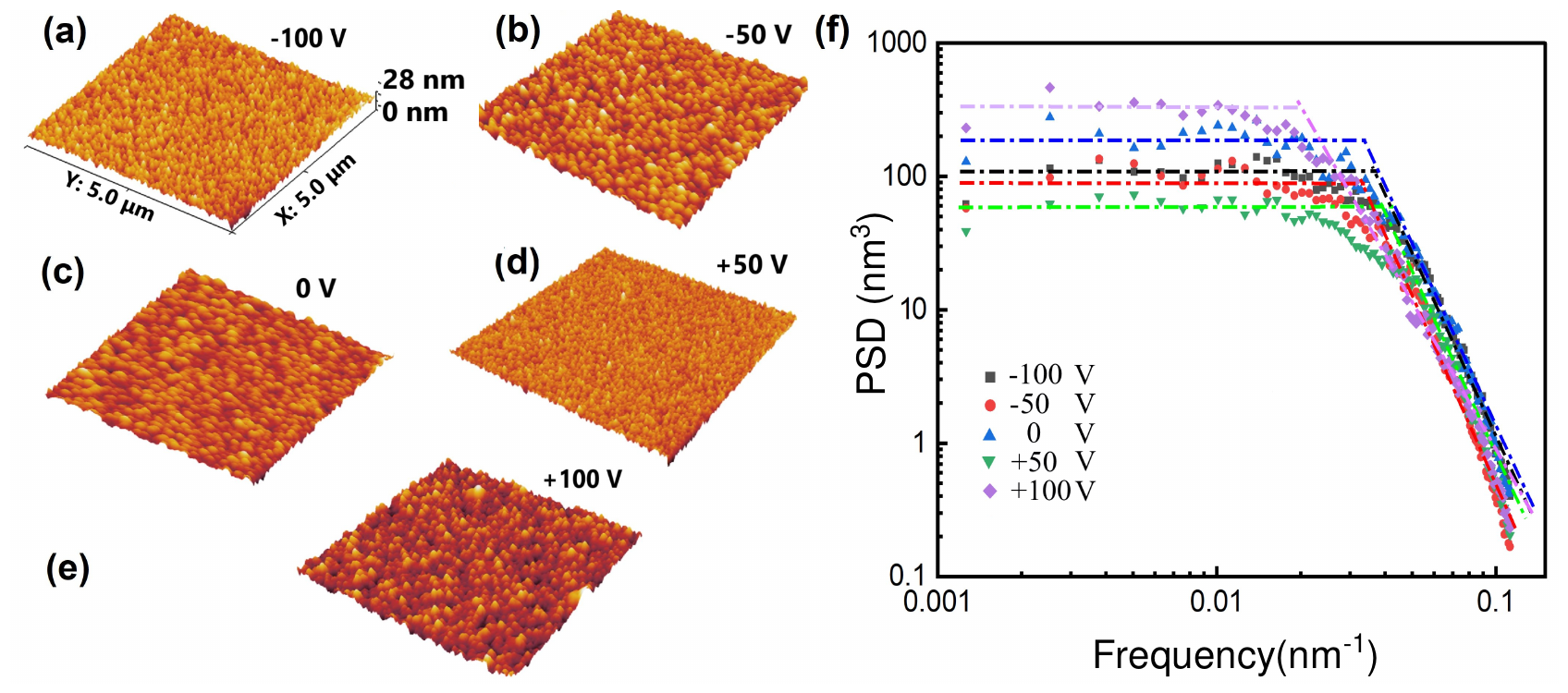} 
  \caption{3D-AFM 5 $\mu$m $\times$ 5 $\mu$m images of  W films at different substrate bias voltages: (a) -100 V, (b) -50 V, (c) 0 V, (d) +50 V, and (e) +100 V. (f) 1D PSD spectra as a function of spatial frequency plotted on a log-log scale. The frequency range is divided into a low-frequency plateau region (indicated by the horizontal dashed line) and a high-frequency linearly varying region ( sloping dashed line). }
  \label{fig:yourlabel}
\end{figure*}

Previous electric transport properties have established that $\alpha$ phase of W has lower resistivity as compared to pure $\beta$ phase \cite{18}. In Figure 2(f), we show the sample thickness at different bias voltages, along with the resistivity of the deposited films, measured using the four-probe method. The film grown at -100 V has the lowest resistivity of around 100 $\mu\Omega$-cm, owing to the larger grain size and $\alpha$ dominated phase observed for this film \cite{35}. At the substrate bias of +50 V, the film has higher resistivity of around 500 $\mu\Omega$-cm, which can be explained due to the $\beta$ dominated tungsten phase as well as smaller grain size that increases the overall scattering of electrons \cite{18,35}. For the film grown at -50 V (570 $\mu\Omega$-cm) as well as at zero bias (1200 $\mu\Omega$-cm, not shown in the Figure 2(f)), the resistivity is found to be higher than the $\beta$ phase due to the semi-crystalline and/or amorphous nature of these films as confirmed by XRD. Finally, the +100 V film has lower resistivity as compared to the film grown at +50 V due the larger grain size as well as the mixed $\alpha$ and $\beta$-W phases observed in this film.

Thus, we have shown that substrate bias can be used to grow desirable $\beta$ phase of W films important for the spintronic application \cite{20,a}. Besides the phase, the surface roughness of the HM plays a crucial role in defining interfacial properties essential for spin Hall Effect or spin-orbit torque-based applications \cite{36,37}. To analyse the effect of substrate bias on the roughness of the grown film, we performed power spectral density (PSD) analysis of the Atomic Force Microscopy (AFM) data shown in Figure 3 for films grown with different substrate biases. Thin films grown by sputtering are generally self-affine surfaces exhibiting statistical scaling properties over different length scales, and PSD analysis helps quantify these properties.  PSD is evaluated from the Fourier transform of the autocorrelation function of power signals across a wide spatial frequency range. In this paper, a one dimensional approach has been adopted for the computation of the PSD function, which follows the equation \cite{38}

\begin{equation}
\text{PSD}(f) = \frac{2\pi}{M_x M_y d} \sum_{j=0}^{M_y-1} |\hat{P}_j(f)|^2
\end{equation}
where \( f \) is the spatial frequency and \( \hat{P}_j(f) \) is the Fourier coefficient of the \( j^{\text{th}} \) line profile defined as:

\begin{equation}
\hat{P}_j(f) = \frac{d}{2\pi} \sum_{n=0}^{M_x-1} z_{nj} e^{-ifnd}
\end{equation}
\( M_x \) and \( M_y \) are the number of points in a profile and number of profiles, respectively. \( d \) is a pixel dimension along the line, and \( z_{nj} \) is the value of the height at the \( n^{\text{th}} \) point in the \( j^{\text{th}} \) line-profile.\\

Figure 3(f) represents the log-log plot of the 1D-PSD profile as a function of spatial frequency for different AFM images shown in Figure 3. All 1D-PSD profiles represent a typical self-affine surface characterised by a low-frequency plateau and a linearly varying high-frequency region\cite{39}. The magnitude of the plateau region in the PSD profile corresponds to the roughness associated with the transition point between the plateau and the linearly varying region, also called the inflection point, which represents the inverse of the correlation length\cite{38,40}.

\begin{figure*}[h]
  \centering
\includegraphics[width=0.8\textwidth]{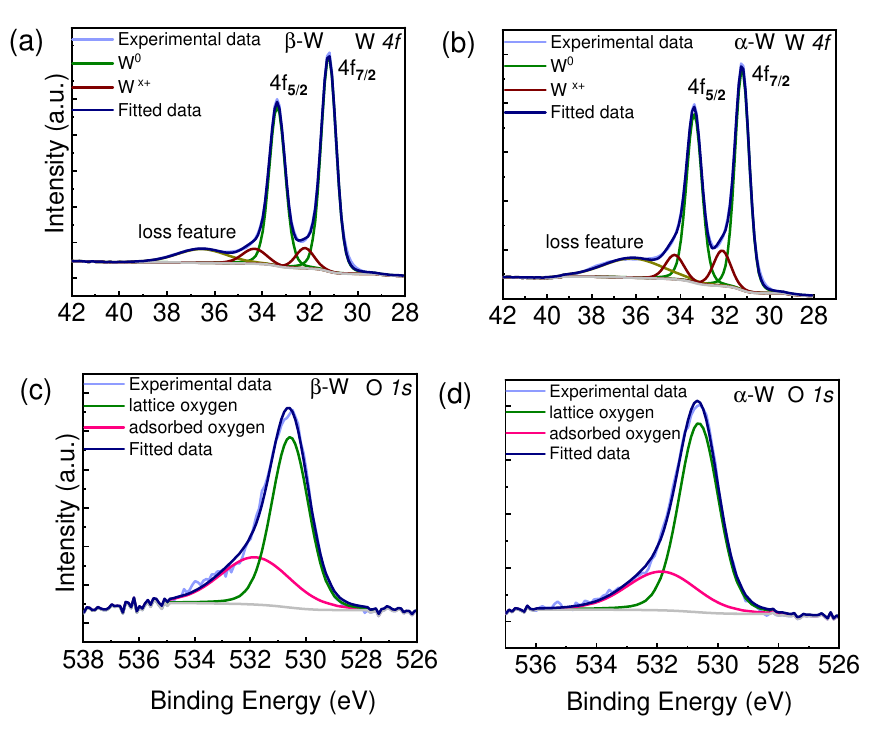} 
  \caption{X-ray photoelectron spectroscopy (XPS) spectra of the W4f and O1s core levels for $\beta$- and $\alpha$- W films. }
  \label{fig:yourlabel}
\end{figure*}

The PSD profile clearly shows that for the film grown with +50 V substrate bias corresponding to the $\beta$-phase of the film, has the lowest plateau region, highlighting the minimum associated roughness. Table 1 shows the roughness obtained using the k-correlation model and the corresponding correlation length obtained from the knee point in the 1D-PSD profile. The lowest correlation length for the +50 V voltage corresponds to the smallest grain size observed in the SEM images.

\begin{table}[htbp]
\centering
\caption{Values of RMS roughness (\( \sigma \)) and Correlation length (\( \tau \)) at different bias voltages}
\setlength{\tabcolsep}{5pt} 
\begin{tabular}{lccccc}
\toprule
\textbf{DC bias voltage (V)} & +100 & +50 & 0 & -50 & -100 \\
\midrule
\textbf{RMS roughness $\sigma$ (nm)} & 0.9 & 0.6 & 1.0 & 0.7 & 0.8 \\
\textbf{Correlation length $\tau$ (nm)} & 53 & 26 & 33 & 30 & 28 \\
\bottomrule
\end{tabular}
\label{tab:my-table}
\end{table}

The XRD, SEM, and AFM analyses confirm that substrate bias effectively controls the phase growth of tungsten (W) films. Specifically, a substrate bias of -100 V promotes the formation of an $\alpha$-dominated W phase, while a +50 V bias favors the $\beta$-dominated W phase, all without significantly affecting the film's roughness. Our results show that the $\beta$-W phase remains stable at higher film thicknesses and can be grown without the need for oxygen plasma. Previous studies have suggested that oxygen inclusion plays a crucial role in determining the phase of the grown tungsten film, with the $\beta$-phase stabilizing at an oxygen concentration in the range of 15-24\%, and the $\alpha$-phase stabilizing at lower oxygen concentrations, around 3-13\% \cite{16,39,aa,bb}. We performed XPS to analyse the chemical composition for the films grown at -100 V and + 50 V bias voltage which corresponds to $\alpha$ and $\beta$-W, respectively. The XPS was performed after etching the surface of the film $\approx 10$ nm using Argon sputtering. In Figure 4, we show the XPS spectra for $\beta$ and $\alpha$-W thin film around W-4f and oxygen (O-1s) region.  For both the films, we observe an asymmetric doublet at 31.1 eV and 33.3 eV attributed to \(4f_{\frac{7}{2}}\) and \(4f_{\frac{5}{2}}\) of metallic tungsten (\(\Delta_{\text{metal}} = 2.2\) eV) \cite{16,39,aa,bb}. Moreover, two distinct peaks at 32.1 eV and 34.3 eV in W 4f spectra indicating a small fraction (0.05) of intermediate oxides state of \(\text{W}^{x+}\), which may originate from the partially oxidized tungsten ions \cite{33,bb}. For both thin films, oxygen concentration was estimated to be around 20\% using the XPS spectra obtained around O-1s region as shown in Figure 4(c) and 4(d). The O-1s peak can be deconvoluted into lattice and adsorbed oxygen with XPS peak around 530.5 eV and 532.1 eV \cite{33,39}. The concentration of lattice oxygen exceeds that of adsorbed oxygen in both $\alpha$ and $\beta$-W films. The residual oxygen in the sputtering chamber acts as the source of oxygen concentration, resulting in both films having almost identical oxygen percentage. Thus, XPS demonstrates that the development of $\beta$ and $\alpha$ phases is exclusively reliant on the bias voltage and is independent of the quantity of oxygen included in the thin film.

\begin{figure}
\centering 
\includegraphics[width=0.5\textwidth,]{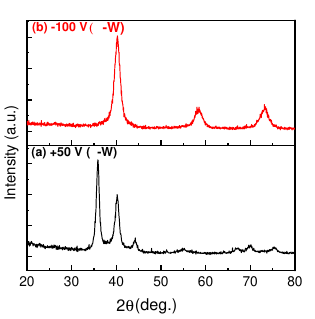}	
\caption{GIXRD pattern of sputter-deposited 30 nm W films at substrate bias voltages of (a) +50 V and (b) -100 V, representing $\beta$- and $\alpha$- phase W films. } 
\label{fig2}
\end{figure}

The above study demonstrates that substrate bias is a powerful tool for selectively controlling the formation of either the $\alpha$ or $\beta$ phase of tungsten, even at higher film thicknesses. To illustrate the potential application for spintronic devices, we grew 30 nm tungsten films on SiO$_2$/Si substrate at substrate biases of -100 V and +50 V, corresponding to the $\alpha$- and $\beta$-phases, respectively. Figure 5 presents the GIXRD patterns for the 30 nm thin films. The film deposited with a +50 V substrate bias is dominated by the $\beta$-W phase, while the film grown at -100 V exhibits a dominance of the $\alpha$-W phase. These findings are consistent with the results shown in the previous section, further confirming the effect of substrate bias on the phase formation of tungsten.

\subsection{\textbf{FMR driven spin pumping in $\boldsymbol{\beta}$-W/Py heterostructure}}

\begin{figure}[h]
     \centering     \includegraphics[width=0.485\textwidth]{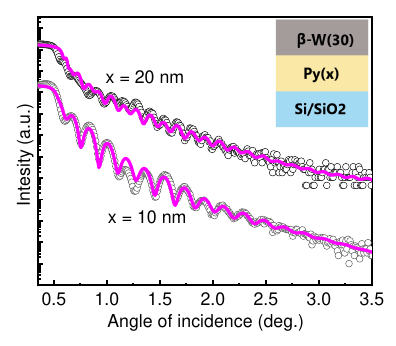}
	\caption{XRR spectra of the $\beta$-W/Py heterostructure. The open symbols represent experimental data, and the dark pink line represents the fit obtained through simulation. The model used for the simulation is shown in the inset.} 
	\label{fig:my_figure}
\end{figure}

We further investigate ferromagnetic resonance (FMR)-driven spin pumping in HM/FM heterostructures \cite{fff,ggg,kkk,lll}. Py was selected as the ferromagnetic (FM) layer, and $\beta$-W was grown using substrate bias to form the $\beta$-W/Py heterostructure. Figure 6 presents the X-ray reflectivity (XRR) spectra, along with simulated curves, for the heterostructure with different FM layer thicknesses (10 nm and 20 nm). The thickness of the $\beta$-W layer is 30 nm. The XRR data exhibit distinct oscillations up to 3$^\circ$, suggesting uniform deposition and a smooth interface, which emphasises that the use of substrate bias preserves the interface quality. Figure 7(a) presents the in-plane MOKE signals for the $\beta$-W/Py heterostructures with varying thicknesses. For comparison, in-plane MOKE signals from $\alpha$-W/Py, grown with a -100V substrate bias, are also shown. In all samples, we observe a square hysteresis loop, indicating in-plane magnetization with low coercivity \cite{aaa,bbb}. The coercivity ($H_c$) value is summarised in Table 2.

We used broadband FMR to investigate spin pumping in the bilayer samples mentioned above, with the external magnetic field ($H_{\text{ext}}$) applied along the film plane. The FMR spectra was obtained over the microwave frequency (\textit{f} ) range of 5–12 GHz. The normalised FMR spectra for $\beta$-W/Py as well as $\alpha$-W/Py samples at fixed frequency of 10 GHz are shown in Figure 7 (b). The FMR spectrum is fitted with sum of the
derivatives of symmetric and antisymmetric Lorentzian functions given below  to obtain the resonance field ($H_r$) and linewidth ($\Delta H$) for all FMR spectra \cite{eee}.

\begin{equation}
\begin{aligned}
\frac{d\chi}{dH} = & -S \frac{\left( \frac{\Delta H}{2} \right)^2 \left( H_{\text{ext}} - H_r \right)}{\left( \left( \frac{\Delta H}{2} \right)^2 + \left( H_{\text{ext}} - H_r \right)^2 \right)^2} \\
& + A \frac{\left( \frac{\Delta H}{2} \right) \left( \left( \frac{\Delta H}{2} \right)^2 - \left( H_{\text{ext}} - H_r \right)^2 \right)}{\left( \left( \frac{\Delta H}{2} \right)^2 + \left( H_{\text{ext}} - H_r \right)^2 \right)^2}
\end{aligned}
\label{eq:3}
\end{equation}

where S and A are the symmetric and antisym-metric absorption coefficients, respectively. Figure 7(c)  shows the variation of $H_r$ as a function of \textit{f} for bilayers samples with  $\beta$  as well as $\alpha$-W. The effective magnetization (\(4\pi M_{\text{eff}}\)) for the films was obtained by fitting the \textit{f} versus $H_r$ plots with Kittel equation \cite{ccc}

\begin{equation}
f = \frac{\gamma}{2\pi} \sqrt{H_r \left(H_r + 4\pi M_{\text{eff}}\right)}
\label{eq:4}
\end{equation}

where $\gamma = 1.85 \times 10^2$ GHz/T is the gyromagnetic ratio. The effective damping constant (\(\alpha_{\text{eff}}\)) and FMR driven spin pumping in bilayer sample can be analysed using the $\Delta H$ of FMR signals. In Figure 7(d) we plot the \textit{f} dependence of $\Delta H$ for all the four bilayer samples and fit the data with the following equation \cite{ddd}: 

\begin{equation}
\Delta H = \frac{4 \pi \alpha_{\text{eff}} f}{\gamma} + \Delta H_0
\label{eq:5}
\end{equation}

where $\Delta H_0$ is the inhomogeneous broadening caused due to sample imperfections. The fitted values of $\alpha_{\text{eff}}$, $4\pi M_\text{eff}$ and $\Delta H_0$ for all the films are summarised in Table 2. 

\begin{table}[htbp]
\centering
\caption{Values of $H_c$, $\alpha_{\text{eff}}$, $4\pi M_\text{eff}$, and $\Delta H_0$ of bilayer structures \cite{sss,ppp,qqq,rrr}}
\setlength{\tabcolsep}{5pt} 
\begin{tabular}{lccccc}
\toprule
Samples & $H_c$(Oe) & $\alpha_{\text{eff}}$ & $4\pi M_\text{eff}$(Oe) & $\Delta H_0$(Oe) \\

\midrule
$\beta$-W(30)/Py(10) & 13  & 0.039 & 3825 & 44 \\
$\beta$-W(30)/Py(20) & 34 & 0.026 & 4219 & 55 \\
$\alpha$-W(30)/Py(10) & 13 & 0.035 & 3380 & 30  \\
$\alpha$-W(30)/Py(20) & 27 & 0.025 & 3930 & 75  \\

\bottomrule
\end{tabular}
\label{tab:my-table}
\end{table}

\begin{figure*}[h]
  \centering  \includegraphics[width=0.9\textwidth]{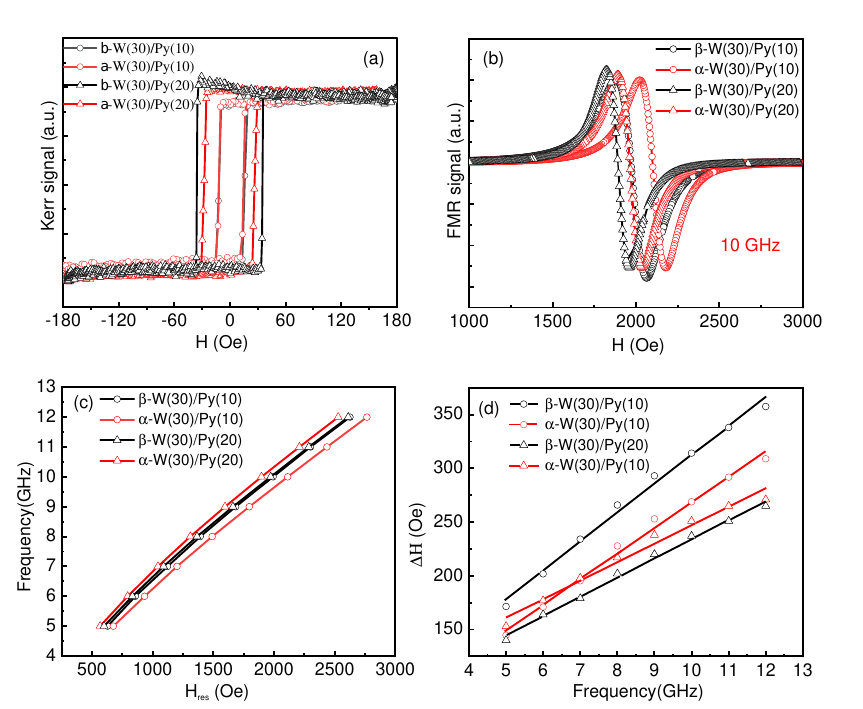} 
  \caption{(a) Room temperature in-plane MOKE signal. (b) Field-swept normalized in-plane FMR spectra recorded at a frequency of 10 GHz. Open symbols indicate experimental data, which are fitted to the derivative of the Lorentzian function as shown in Equation (3). (c) Variation of $H_r$ with frequency. Open symbols represent the experimental data, and the fit follows the Kittel equation (4). (d) Variation of $\Delta H$ with frequency. Open symbols represent the experimental data, and the fit follows the linewidth equation (5).}
  \label{fig:yourlabel}
\end{figure*}

 $\Delta H_0$ varies from 30 Oe to 75 Oe for all the samples, suggesting good quality of the thin grown bilayer samples, which was also confirmed by the observation of well defined oscillation in X-ray reflectivity data shown in Figure 6. The values of $4\pi M_\text{eff}$ regardless of the $\alpha$ or $\beta$-W phase, remained almost the same, suggesting that the W layer deposition conditions had no significant effect. However, we observe a significant change in \(\alpha_{\text{eff}}\) as Py layer thickness is varied from 10 nm to 20 nm. For $\beta$-W/Py(20 nm) and $\alpha$-W/Py(20 nm) films,  \(\alpha_{\text{eff}}\) was found to be almost similar with values of 0.026 and 0.025, respectively. As we decrease the FM layer thickness to 10 nm while keeping the HM layer constant, the \(\alpha_{\text{eff}}\) for $\beta$-W/Py(10 nm) and $\alpha$-W/Py(10 nm) was found to be 0.039 and 0.035, respectively. The enhancement in  \(\alpha_{\text{eff}}\) with decreasing FM layer thickness is a signature for spin pumping.  Moreover, higher value of  \(\alpha_{\text{eff}}\) for $\beta$-W/Py(10 nm) as compared to  $\alpha$-W/Py(10 nm) indicates a more efficient transfer of spin angular momentum in $\beta$-W, primarily due to its stronger spin-orbit coupling (SOC) \cite{mmm}. We further quantify the spin pumping efficiency in terms of effective spin mixing conductance (\( g_{\text{eff}}^{\uparrow\downarrow} \)). In our case, the thickness of W (in both phases) is fixed at 30 nm, which is much greater than the spin diffusion length of W (around 4 nm). This allows us to neglect the effect of spin backflow on the spin pumping efficiency. The \( g_{\text{eff}}^{\uparrow\downarrow} \) can be determined by plotting \(\alpha_{\text{eff}}\) against the thickness of the ferromagnetic layer using the following formula: \cite{fff,ggg,hhh,iii,jjj}

\begin{equation}
\alpha_{\text{eff}} = \alpha_0 + g \mu_B \frac{g^{\uparrow \downarrow}}{4 \pi M_{\text{eff}}} \frac{1}{t_{\text{FM}}}
\label{eq:6}
\end{equation}

where $\alpha_0$ is the intrinsic Gilbert damping constant, $t_{\text{FM}}$ is the thickness of the ferromagnetic layer. From Equation (6), we observe that the damping constant depends linearly on the inverse of the ferromagnetic thickness. This behavior is evident in both of our batches: batch 1 with $\alpha$-W(30 nm)/Py($x$)/SiO$_2$/Si and batch 2 with $\beta$-W(30 nm)/Py($x$)/SiO$_2$/Si. In both cases, \(\alpha_{\text{eff}}\) decreases with increasing Py thickness. The linear fit of our data using Equation (6) (not shown here) yields $\alpha_0 = 0.015$ [0.013] for $\alpha$-W(30 nm)/Py($x$) [$\beta$-W(30 nm)/Py ($x$)], where $x = 10$ nm, $20$ nm. The  \( g_{\text{eff}}^{\uparrow\downarrow} \) is calculated using the slope \( \frac{g \mu_B g^{\uparrow\downarrow}}{4\pi M_{\text{eff}}} \). The obtained values of \( g_{\text{eff}}^{\uparrow\downarrow} \) are $4.5 \, (\pm 0.5) \, \text{nm}^{-2}$ and $3.0 \, (\pm 0.2) \, \text{nm}^{-2}$ for $\beta$-W/Py and $\alpha$-W/Py, respectively \cite{nnn,ooo}. Therefore, we observe that \( g_{\text{eff}}^{\uparrow\downarrow} \) strongly depends on the crystal structure of the W phase and is larger for $\beta$-W/Py \cite{nnn}. The spin pumping efficiency is enhanced in $\beta$-W, which might be due to the different atomic and electronic arrangements of the W-phases at the interface. These differences ultimately modify the interfacial spin-orbit coupling (SOC), thereby tuning the spin pumping efficiency.

\section{Summary and conclusions}

In summary, we have systematically investigated the substrate bias-dependent phase transition of tungsten  films without using oxygen plasma and its impact on spin pumping efficiency in W/Py heterostructures. Our results demonstrate that substrate bias plays a role in controlling the energy of deposited atoms, thereby influencing the properties of the thin films produced. The $\beta$-W exhibited higher \(\alpha_{\text{eff}}\) compared to $\alpha$-W. The \( g_{\text{eff}}^{\uparrow\downarrow} \) , which characterizes the interface, was also higher in $\beta$-W/Py than in $\alpha$-W/Py. This indicates that spin transfer in $\beta$-W/Py is  larger than in $\alpha$-W/Py, primarily due to the interface being engineered differently by different bias voltage.  Thus, this study demonstrates that substrate bias not only provides precise control over the phase formation of tungsten ($\alpha$ and $\beta$) but also preserves interface quality, making it an essential tool for engineering thin-film properties. The superior spin transfer efficiency of $\beta$-W underscores its potential for advanced spintronic applications.

\section*{CRediT authorship contribution statement}

Abhay Singh Rajawat: Conceptualization, Methodology, Data curation, Writing-original draft. Naim Ahmad: Data curation, Formal Analysis. Risvana Nasril: Data curation, Software. Tasneem Sheikh: Data curation, Software. Mohammad Muhiuddin: Data curation, Formal Anlalysis. Savita Sahu: Data curation. Ashwani Gautam: Data curation, Software. Akumar: Formal Analysis. Md. Imetyaz Ahmad: Formal Analysis, Writing-review \& editing. G.A. Basheed: Resources, Formal Analysis, Writing-review \& editing. Mohammad R.Rahman: Resources, Formal Analysis, Writing-review \& editing. Waseem Akhatar: Conceptualization, Supervision, Validation, Resources, Writing-review \& editing, Project administration.

\section*{Declaration of Competing Interest}
The authors declare that they have no known competing financial
interests or personal relationships that could have appeared to influence the work reported in this paper.

\section*{Acknowledgements}

This work was financially supported by Science and Engineering Research Board (SERB) Start-Up Research Grant (SRG/2021/002186) and University Grants Commission (UGC) Start-up research grant (F.30-569/2021). The authors gratefully acknowledge the Central Research Facility, National Institute of Technology, Karnataka for the XRR measurement.


\bibliographystyle{elsarticle-num} 
\bibliography{ref}

\end{document}